\documentclass[aps,twocolumn,showpacs,superscriptaddress]{revtex4-1}

\usepackage{amssymb}
\usepackage{amsmath}
\usepackage{graphicx}
\usepackage{hyperref}
\usepackage{siunitx}
\hypersetup{colorlinks=true,
            linkcolor=blue,
            citecolor=blue,
            urlcolor=orange}
        
\usepackage[dvipsnames]{xcolor}

\begin{document}

\title{Commensurate and {spiral} magnetic order in the doped \\two-dimensional Hubbard model: dynamical mean-field theory analysis}
\author{I. A. Goremykin}
\affiliation{Center for Photonics and 2D Materials, Moscow Institute of Physics and Technology, Institutsky lane 9, Dolgoprudny
141700, Moscow region, Russia}
\author{A. A. Katanin}
\affiliation{Center for Photonics and 2D Materials, Moscow Institute of Physics and Technology, Institutsky lane 9, Dolgoprudny
141700, Moscow region, Russia}
\affiliation{M. N. Mikheev Institute of Metal Physics, Kovalevskaya Street 18,
Ekaterinburg {620219}, Russia.}

\begin{abstract}
We develop {a} dynamical mean-field theory approach for the {spiral} magnetic order,
{changing} to a local coordinate frame with preferable spin alignment along {the} $z$-axis, which can be considered with the impurity solvers treating {the} {spin diagonal local Green's function. We furthermore solve} the Bethe-Salpeter equations for nonuniform dynamic magnetic susceptibilities in the local coordinate frame. We apply this approach to describe the evolution of magnetic order with doping in the $t-t'$ Hubbard model with $t'=0.15$, which is appropriate for {the} description of the doped La$_2$CuO$_4$ high-temperature superconductor. We find that with doping the antiferromagnetic order changes to {the} $(Q,\pi)$ incommensurate one, and then to the paramagnetic phase. {The spectral weight at the Fermi level is suppressed near half filling and continuously increases with doping.} The dispersion of  holes in {the} antiferromagnetic phase shows qualitative agreement with the results of the $t$-$J$ model consideration.
In the incommensurate phase we find two branches of hole dispersions, one of {which} {crosses} the Fermi level. The resulting Fermi surface forms hole pockets. We also  consider {the} dispersion of the magnetic excitations, obtained from the non-local dynamic magnetic susceptibilities.
The transverse spin excitations are gapless, fulfilling {the} Goldstone theorem; 
{in contrast to the mean-field approach the obtained magnetic state is found to be stable.}
The longitudinal excitations are characterized by a small gap, showing {the} rigidity of {the} spin excitations. For realistic hopping and interaction {parameters} we reproduce {the} experimentally measured spin-wave dispersion of La$_2$CuO$_4$.
\end{abstract}

\maketitle

\section{Introduction}

The properties of {a} two-dimensional antiferromagnet doped by holes became one of the central interests in condensed-matter physics starting {in} the high-$T_c$ era. Despite {the fact} that the commensurate long-range magnetic order is quickly destroyed by doping (see, e.g., Refs. \cite{Plakida,StaticOrd}), the short-range magnetic order 
is present in high-$T_c$ compounds.  The observed short-range magnetic order is incommensurate with {the} {preferred} wave vector ${\bf Q}=(\pi,\pi-\delta)$ and small incommensurability parameter $\delta$ \cite{ExpQ1,ExpQ2,ExpQ3,StaticOrd,Plakida}.  This short-range order is considered {to be} one of the viable scenarios {for} pseudogap formation \cite{PseudogapSF,PseudogapSad,PseudogapOnufr,PseudogapToschi,PseudogapMag1,PaseudogapMag2,PseudogapMag3,Sachdev1,Sachdev2}. The incommensurate magnetic order is also observed on frustrated lattices, such as the triangular lattice \cite{TLAF1,TLAF2}. 

Although the long-range magnetic order is absent at substantial doping in high-$T_c$ compounds, studying {the} properties of long-range commensurate and incommensurate magnetic ordered states in the two-dimensional Hubbard model represents a certain interest since such order models properties of the paramagnetic phase with pronounced magnetic correlations. {As a result,} mean-field studies were performed to investigate the commensurate \cite{SDW1,SDW2} and incommensurate {spiral} \cite{SDWIc1,SDWIc2,SDW3,SDWIc1,SDWIc2,SDWIc3,SDW4,SDWOur,BM} magnetic orders in the doped square lattice Hubbard model. Incommensurate magnetic order on the same lattice was later studied within the slave-boson approach in Ref. \cite{SBIncomm}. The latter studies \cite{SDW3,SDW4,SDWOur,SBIncomm} {showed}, however, that the incommensurate magnetically ordered states in the hole doped Hubbard model are thermodynamically unstable within the mean-field approach, which yields a phase separation \cite{SDWOur,SBIncomm} of incommensurate magnetic order into domains with incommensurate and commensurate magnetic states. However, the above-discussed mean-field approaches do not consider (or consider only approximately for the slave-boson approach) {the} effect of local electronic correlations, which are important for sufficiently strong Coulomb repulsion. 

The dynamical mean-field theory (DMFT) approach offers {the} possibility of studying commensurate \cite{DMFT} and incommensurate {spiral} \cite{DMFT_Incomm_Licht,DMFT_Incomm} magnetic orders {while} taking into account local magnetic correlations. While the commensurate magnetic order within {the} DMFT approach for the square lattice Hubbard model {is} also shown to suffer from phase separation \cite{Pruschke}, {a} detailed analysis of {the} possibility of incommensurate magnetic order in the doped Hubbard model on the square lattice was not performed, to our knowledge, {except for the case of {only} nearest-{neighbor} hopping \cite{DMFT_Incomm_Licht}}. Moreover, to analyze the stability of incommensurate magnetic order, nonlocal dynamic magnetic susceptibilities should be considered. While {a} general formalism for calculating such susceptibilities was recently proposed \cite{BS_Toschi}, its application to {the} incommensurate magnetically ordered phase is challenging.

In the present paper we reformulate the dynamical mean-field theory approach for treatment of an incommensurate {spiral} magnetic order in the Hubbard model in {a} way that allows using impurity solvers {which} treat {spin diagonal local Green's functions}. We furthermore apply the formalism of Ref. \cite{BS_Toschi} {to treat} nonlocal magnetic static and dynamic susceptibilities. We consider {the} square lattice Hubbard model with nearest- and next-nearest-neighbor hopping and calculate the sublattice {magnetization and} incommensurability parameter as a function of doping and obtain longitudinal and transverse magnetic excitation spectra.

The plan of the paper is {as follows}. In Sec. II we formulate the model and present the method {to} treat commensurate and incommensurate ordered states. In Sec. III we present the results for the Fermi surfaces, {the} hole dispersion, and the dispersion of longitudinal and transverse magnetic excitations, obtained from the dynamic spin susceptibility. In Sec. IV we present our conclusions. {In {the} Appendix we discuss the relation of the considered approach to the static mean-field theory}.

\vspace{-0.1cm}
\section{Model and method} 

We consider the Hubbard model on the square lattice 
\begin{equation} 
    H = 
    - \sum_{
    {i, j, \sigma
    }} 
t_{ij}{c_{i\sigma}^\dagger c_{j\sigma}}+U\sum_i{n_{i\uparrow}n_{i\downarrow}},
\label{H}
\end{equation}
with hopping $t_{ij}=t$ between nearest neighbors (which is used as a unit of energy) and $t_{ij}=-t'$ for next-nearest neighbors. We consider {spiral} spin density wave magnetic order with wave vector ${\bf Q}$, which is, in general, incommensurate. 

\vspace{-0.45cm}
 \subsection{Derivation of DMFT equations in the {spiral} case}
 
 To study the model (\ref{H}) we apply the DMFT approach \cite{DMFT}. For the commensurate case ${\bf Q}=(\pi,\pi)$ the DMFT equations are standard (see Ref. \cite{DMFT}). To study incommensurate {spiral order} we perform rotation of {the} coordinate system in spin space by an angle $\theta=\mathbf{QR}_{i}$ to {the} local coordination frame, in which the preferred direction of the spin alignment is along the $z$ direction. This is different from the approach of Refs. \cite{DMFT_Incomm_Licht,DMFT_Incomm}, where the spins were aligned along {the} $x$ direction. {{The} advantage of {having} the spin alignment along the $z$ direction lies in its codirection
 with the spin-quantization axes. In this case {the} electron Green's functions become spin diagonal in the local coordinate frame. This allows us to use the segment version of {the continuous-time quantum Monte Carlo} (CT-QMC) solver {to treat} density-density interactions
 (see below). }

Let us consider a spiral spin density wave with spins rotating in the $xz$ plane with wave vector $\mathbf{Q}.$ {We consider operators $d_{i\sigma},d^+_{i\sigma}$ in a local coordinate system
where all spins are aligned along $z$ axis. To this end we perform the unitary transformation}%
 \begin{equation}
\left(
\begin{tabular}
[c]{l}%
$d_{i\uparrow}$\\
$d_{i\downarrow}$%
\end{tabular}
\right)  =\left(
\begin{tabular}
[c]{ll}%
$\cos(\theta/2)$ & $\sin(\theta/2)$\\
$-\sin(\theta/2)$ & $\cos(\theta/2)$%
\end{tabular}
\ \right)  \left(
\begin{tabular}
[c]{l}%
$c_{i\uparrow}$\\
$c_{i\downarrow}$%
\end{tabular}
\right).
\end{equation}
The on-site Hubbard interaction is $SU(2)$
invariant,%
\begin{align*}
n_{i\uparrow}n_{i\downarrow}   =
d_{i\uparrow}^{+}d_{i\uparrow}d_{i\downarrow}^{+}d_{i\downarrow}.
\end{align*}
Therefore, the local problem is formulated straightforwardly,%
\vspace{-0.2cm}
\begin{align}
S_{\text{loc}}&=-T\sum\limits_{\sigma}%
{\displaystyle\int_{0}^{\beta}}
d\tau%
{\displaystyle\int_{0}^{\beta}}
d\tau^{\prime} {\zeta^{-1}_{\sigma}(\tau-\tau^{\prime})d_{i\sigma}^{+}d_{i\sigma}}\notag \\
&+U%
{\displaystyle\int_{0}^{\beta}}
d\tau d_{i\uparrow}^{+}d_{i\uparrow}d_{i\downarrow}^{+}d_{i\downarrow},
\label{local_problem}
\end{align}
where $\zeta_{\sigma}(\tau-\tau^{\prime})$ is the bath Green's function.

Let us now consider the nonlocal part. We {consider} Fourier transformed
operators
\begin{align}
c_{\mathbf{k}\sigma}=\sum\limits_{\sigma'}(\mathcal{M}^+_{\sigma\sigma^{\prime}}d_{\mathbf{k+Q}/2\mathbf{,}%
\sigma^{\prime}}+\mathcal{M}^-_{\sigma\sigma^{\prime}}d_{\mathbf{k-Q}%
/2\mathbf{,}\sigma^{\prime}})/2,
\end{align}
where 
\vspace{-0.4cm}
\begin{equation}
\mathcal{M}^\pm=\left(
\begin{array}
[c]{cc}%
1 & \pm i\\
\mp i & 1
\end{array}
\right)  =\sigma^{0}\mp \sigma^{y},%
\end{equation}
and $\sigma^a$ are the Pauli matrices.
Therefore,%
\begin{align}
c_{\mathbf{k}\uparrow}^{+}c_{\mathbf{k}\uparrow}+c_{\mathbf{k}\downarrow}%
^{+}c_{\mathbf{k}\downarrow} &  =\sum_{\alpha\sigma\sigma'} (d_{\mathbf{k+\alpha Q}/2\mathbf{,}%
\sigma}^{+} \mathcal{M}_{\sigma \sigma'}^\alpha d_{\mathbf{k+\alpha Q}/2\mathbf{,}\sigma'}^{{}})/2 \notag \\
&=\sum_\alpha D_{\mathbf{k+}\alpha\mathbf{Q}/2\mathbf{,}\alpha}^{+}D_{\mathbf{k+}%
\alpha\mathbf{Q}/2\mathbf{,}\alpha},
\end{align}
where $\alpha=\pm1$ is the band index and we have introduced the operators%
\vspace{-0.2cm}
\begin{equation}
D_{\mathbf{k,}\pm}=\left(  d_{\mathbf{k}\uparrow}\pm id_{\mathbf{k}%
\downarrow}\right)  /\sqrt{2}.
\end{equation}
The kinetic term then reads%
\begin{align}
\sum_{\mathbf{k\sigma}}\epsilon_{\mathbf{k}}c_{\mathbf{k}\sigma}%
^{+}c_{\mathbf{k}\sigma}&=
\sum_{\mathbf{k}\alpha}\epsilon
_{\mathbf{k-}\alpha\mathbf{Q}/2}D_{\mathbf{k}\alpha}^{+}D_{\mathbf{k}\alpha},
\end{align}
where $\epsilon_{\bf k}=-2t(\cos k_x+\cos k_y)+4t' \cos k_x \cos k_y$. Denoting the self-energy in the rotated frame $d^+_{i\sigma}$ {and} $d_{i\sigma}$ as $\Sigma_\sigma(i\nu)$, we write 
the lattice Green's function 
in the form
\begin{widetext}
\vspace{-0.25cm}
\begin{equation}
    G_{\mathbf{k}}^{\sigma \sigma'}(i\nu) = -\langle d_{\mathbf{k}\sigma}(\tau)d_{\mathbf{k}%
\sigma'}^{+}(0)\rangle_{i\nu}=
    \begin{pmatrix}
    i\nu + \mu - \Sigma_{\uparrow} - \frac{\epsilon_{\mathbf{k-Q}/2} + \epsilon_{\mathbf{k+Q}/2}}{2} &
    \frac{\epsilon_{\mathbf{k-Q}/2} - \epsilon_{\mathbf{k+Q}/2}}{2i} \\
    - \frac{\epsilon_{\mathbf{k-Q}/2} - \epsilon_{\mathbf{k+Q}/2}}{2i} & i\nu + \mu - \Sigma_{\downarrow} - \frac{\epsilon_{\mathbf{k-Q}/2} + \epsilon_{\mathbf{k+Q}/2}}{2}
    \end{pmatrix}^{-1}.
    \label{GFss1}
\end{equation}
\vspace{-0.27cm}
In the explicit form
\begin{align}
G_{\mathbf{k}}^{\sigma\sigma}(i\nu)
=\frac{i\nu+\mu-(\epsilon_{\mathbf{k-Q}/2}%
+\epsilon_{\mathbf{k+Q}/2})/2-\Sigma_{-\sigma}(i\nu)}{(\phi_{\nu}-\epsilon_{\mathbf{k-Q/}2})(\phi_{\nu}-\epsilon_{\mathbf{k+Q/}2})-\left[\Sigma_{\uparrow}(i\nu
)-\Sigma_{\downarrow}(i\nu)\right]^{2}/4},\label{Gss}\\
G_{\mathbf{k}}^{\sigma,-\sigma}(i\nu)
=\frac{{i\sigma}(\epsilon_{\mathbf{k{-}Q}/2}%
-\epsilon_{\mathbf{k{+}Q}/2})/2}{(\phi_{\nu}-\epsilon_{\mathbf{k-Q/}2})(\phi_{\nu}-\epsilon_{\mathbf{k+Q/}2})-\left[\Sigma_{\uparrow}(i\nu)-\Sigma_{\downarrow}%
(i\nu)\right]^{2}/4},
\label{Gsms}
\end{align}
\end{widetext}
where $\phi_{\nu}=i\nu+\mu-\left[\Sigma_{\uparrow}(i\nu)+\Sigma_{\downarrow}(i\nu)\right]/2$. 

The self-consistent equation reads%
\begin{equation}
G_{\rm loc}^{\sigma}(i\nu)\equiv\frac{1}{\zeta_{\sigma}^{-1}%
(i\nu)-\Sigma_{\sigma}(i\nu)}=\sum_{\mathbf{k}}G_{\mathbf{k}}^{\sigma\sigma}(i\nu).
\label{self_con_eqv}
\end{equation}
As we discuss in {the} Appendix, in the mean-field approximation for the Anderson impurity model $\Sigma_\sigma=U n_{-\sigma}$, Eq. (\ref{self_con_eqv}) allows {us} to reproduce the standard mean-field approach for {the} incommensurate spin density wave order. Note also that for the commensurate case the Green's functions $G_{\mathbf{k}%
}^{\sigma\sigma}(i\nu)$ and $G_{\mathbf{k}}^{\sigma,-\sigma}(i\nu)$ correspond to the
intra- and intersublattice Green's functions in the approaches, which split the lattice into two sublattices. This establishes the relation between the
DMFT approaches for the commensurate and incommensurate cases.

{The corresponding lattice Green's functions for ${\bf Q}\neq {\bf 0}$ read}{
\begin{align}
    \mathfrak{G}_{\mathbf{k}\sigma}(i\nu)&=-\langle T {c}_{\mathbf{k}\sigma} (\tau) {c}_{\mathbf{k}\sigma}^{+}(0)\rangle_{i\nu}  \label{Gcc}\\&=
   \frac{1}{4} \sum_{\sigma',\alpha=\pm 1} \left[ G_{\mathbf{k}+\alpha \mathbf{Q}/2}^{\sigma' \sigma'}(i\nu) - i \alpha {\sigma'}G_{\mathbf{k}+\alpha \mathbf{Q}/2}^{\sigma', -\sigma'}(i\nu)\right].\notag 
\end{align}}
Note that the resulting Green's function does not depend on spin projection.

\subsection{Relation of the two-particle quantities in the global and local reference frame\label{Rel2P}}

For the {two}-particle quantities of {$c$} operators we find the following representation in terms of {$d$} operators:
\begin{align}
\sum_\mathbf{k} c_{\mathbf{k}}^{+}\sigma^{0,y}c_{\mathbf{k+q}}  &
=\sum_\mathbf{k}d_{\mathbf{k}}^{+}\sigma^{0,y}d_{\mathbf{k}%
+\mathbf{q}},\\
\sum_\mathbf{k} c_{\mathbf{k}}^{+}\sigma^{\pm}c_{\mathbf{k+q}%
}&=\sum_\mathbf{k} d_{\mathbf{k
}}^{+}\sigma^\pm d_{\mathbf{k\pm Q}
+\mathbf{q}},%
\end{align}
where $\sigma^{\pm}=\sigma^{z}\pm i\sigma^{x}$ are the in-plane spin raising and lowering matrices.

For the out-of-plane spin component $y$ and the density correlators we therefore have
\begin{align}
&\sum_{\mathbf{k}, \mathbf{k}'} \langle\langle c_{\mathbf{k}}^+ \sigma^{0,y} c_{\mathbf{k}+\mathbf{q}}| c_{\mathbf{k}'}^+ \sigma^{0,y} c_{\mathbf{k}'-\mathbf{q}} \rangle\rangle \notag \\&= 
\sum_{\mathbf{k}, \mathbf{k}'} \langle\langle d_{\mathbf{k}}^+ \sigma^{0,y} d_{\mathbf{k}+\mathbf{q}}| d_{\mathbf{k}'}^+ \sigma^{0,y} d_{\mathbf{k}'-\mathbf{q}} \rangle\rangle,
\end{align}
while for the in-plane spin components we have
\begin{align}
&\sum_\mathbf{k,k'} \langle\langle c_{\mathbf{k}}^{+}\sigma^{a}c_{\mathbf{k+q}}|c_{\mathbf{k}%
^{\prime}}^{+}\sigma^{b}c_{\mathbf{k}^{\prime}\mathbf{-q}^{\prime}}%
\rangle\rangle \\
&  =\sum_\mathbf{k,k'}\langle\langle d_{\mathbf{k}}^{+}\sigma^a d_{\mathbf{k}+a \mathbf{Q}+\mathbf{q}}|d_{\mathbf{k'}}^{+}%
\sigma^b d_{\mathbf{k'}-a \mathbf{Q}-\mathbf{q}}%
\rangle\rangle \delta_{\mathbf{q}+a \mathbf{Q},\mathbf{q'}-b \mathbf{Q}},\notag
\end{align}
where $a,b=\pm$. We note that the in-plane and out-of-plane (or density) excitations are generally coupled in the local coordinate frame (see {the} next section), which {also} yields coupling of these excitations in the global coordinate frame via the nonzero correlators $ \langle\langle c_{\mathbf{k}}^{+} \sigma^{0,y} c_{\mathbf{k}+\mathbf{q}} | c_{\mathbf{k}'}^{+} \sigma^{\pm} c_{\mathbf{k}'-\mathbf{q'}} \rangle\rangle$ at ${\mathbf q}'={\mathbf q}\pm \mathbf{Q}$, (cf. Ref. \cite{SDW3}). At the same time, the correlators in the local coordinate frame are diagonal in momentum, which makes this frame somewhat more advantageous {than} the standard global coordinate frame considered in Ref. \cite{SDW3}.

In the commensurate case ($2{\mathbf Q}\equiv0$) we find
\begin{align}
    &\sum_{\mathbf{k}, \mathbf{k}'} \langle\langle c_{\mathbf{k}}^{+} \sigma^{x,z} c_{\mathbf{k}+\mathbf{q}} | c_{\mathbf{k}'}^{+} \sigma^{x,z} c_{\mathbf{k}'-\mathbf{q}} \rangle\rangle \\&= \sum_{\mathbf{k}, \mathbf{k}'}  
     \langle\langle d_{\mathbf{k}}^{+} \sigma^{x,z} d_{\mathbf{k}+\mathbf{q}+\mathbf{Q} } | d_{\mathbf{k}' }^{+}\sigma^{x,z} d_{\mathbf{k}'-\mathbf{q} -\mathbf{Q}} \rangle\rangle, \notag  
\end{align}
which corresponds to the decoupling of longitudinal and transverse in-plane excitations in this case (see also {the} next section).

\subsection{Calculation of the nonlocal susceptibilities in the local reference frame}

To study the two-particle properties we perform ladder summation of the diagrams for the non-uniform dynamic susceptibility in the particle-hole channel
(cf. Refs. \cite{MyEDMFT,OurRev,My_BS})
\begin{align}
\mathcal{\chi}_{q}^{\mathfrak{m}\mathfrak{m'}}&=
\sum_{k,k'}\langle\langle{d^+_{k,\sigma} d_{k+q,\sigma'}|d^+_{k'+q,\sigma'''} d_{k',\sigma''}}\rangle\rangle \\
&= \sum_{\nu,\nu'}\left[  (\chi_{q,\nu}^{0,\mathfrak{m}\mathfrak{m'}})_{\mathfrak{m}\mathfrak{m'}}^{-1} \delta_{\nu\nu^{\prime}}\right.\label{chi_BS}-\left.
\Phi_{\omega,\nu\nu^{\prime}}
^{\mathfrak{m}\mathfrak{m'}}\right]  _{\mu
\mu^{\prime}}^{-1},\notag
\end{align}
where we introduce composite indexes $q = (\mathbf{q},i\omega)$, $\mu=(\nu,\mathfrak{ m})$, $\mu'=(\nu',\mathfrak{ m}')$, $\mathfrak{ m}=(\sigma,\sigma')$ and $\mathfrak{ m'}=(\sigma'',\sigma''')$. {This ladder summation is similar to that considered previously for the paramagnetic \cite{DMFT,OurRev,MyEDMFT,My_BS,OurSC} and ferro- and antiferromagnetic \cite{BS_Toschi} phases, and allows us to obtain the non-uniform dynamic susceptibilities within the DMFT approach, which respects conservation laws (see, e.g., Refs. \cite{OurSC,OurRev,KrienThes})}. The bare susceptibility is
\begin{equation}
\chi^{0,(\sigma\sigma'),(\sigma''\sigma''')}_{ q,\nu}=-T\sum\limits_{\mathbf{k}}G^{\sigma'' \sigma}_{k}G^{\sigma'\sigma'''}_{k+q}\label{chi0},
\end{equation}
where we use the {three}-vector notations $k=({\bf k},i\nu)$ in the nonlocal Green's function  $G_k^{\sigma\sigma'}=G_{\mathbf k}^{\sigma\sigma'}(i\nu)$ [see Eqs. (\ref{GFss1})--(\ref{Gsms})].
{Quantities with composite spin indexes $\mathfrak{m}$ {and} $\mathfrak{m'}$ can be viewed as 4$\times$4 matrices with the indexes labeled in the order $\uparrow\uparrow$, $\uparrow\downarrow$, $\downarrow\uparrow$, and $\downarrow\downarrow$.}
{The matrix-valued local irreducible vertex $\Phi$ then takes the form}
\begin{equation}
  {\hat \Phi}_{\omega, \nu\nu'} = \left(
  \begin{matrix}
    {\Phi^{||,\uparrow \uparrow}_{\omega, \nu\nu'}} & 0 & 0 & {\Phi^{||,\uparrow \downarrow}_{\omega, \nu\nu'}}\\
    0 & {\Phi^{\bot,\uparrow \downarrow}_{\omega, \nu\nu'}} & 0 & 0 \\
    0 & 0 & {\Phi^{\bot,\downarrow \uparrow}_{\omega, \nu \nu'}} & 0 \\
    {\Phi^{||,\downarrow \uparrow}_{\omega, \nu\nu'}} & 0 & 0 & {\Phi^{||,\downarrow \downarrow}_{\omega, \nu\nu'}}
  \end{matrix}
  \right),
\end{equation}
where the components $\Phi^{||,\bot}$  in longitudinal and transverse channels are evaluated from local vertices $\Gamma^{||,\bot}$ in {the} corresponding channel via the Bethe-Salpeter equations
\begin{align}
{\Gamma^{||,{\sigma\sigma'}}_{\omega,\nu \nu'}}&=\left[  ({\Phi^{||,{\sigma\sigma'}}_{\omega,\nu \nu' }})_{\lambda\lambda'}^{-1} \right.-\left.
 {\chi^{0,||,\sigma\sigma'}_{\omega,\nu}} \delta_{\nu\nu^{\prime}}\right]  _{\lambda\lambda^{\prime}}^{-1},\\
 {\Gamma^{\bot,{\sigma,-\sigma}}_{\omega,\nu \nu'}}&=\left[  ({\Phi^{\bot,{\sigma,-\sigma}}_{\omega,\nu \nu' }})^{-1} \right.-\left.
 {\chi^{0,\bot,\sigma,-\sigma}_{\omega,\nu}} \delta_{\nu\nu^{\prime}}\right]  _{\nu \nu^{\prime}}^{-1},\label{local_BS}
\end{align}
{{where} composite indexes {$\lambda=(\sigma,\nu)$ {and} $\lambda'=(\sigma',\nu')$, local bare susceptibilities} are ${\chi^{0,||,\sigma\sigma'}_{\omega,\nu}}=-T {G_{\rm loc}^{\sigma}}(i \nu) {G_{\rm loc}^{\sigma}}(i \nu+i \omega) \delta_{\sigma\sigma'}$ and ${\chi^{0,\bot,\sigma\sigma'}_{\omega,\nu}}=-T {G_{\rm loc}^{\sigma}}(i \nu) {G_{\rm loc}^{\sigma'}}(i \nu+i \omega)$.} 
{Finally, the local vertices $\Gamma^{||(\bot),{\sigma\sigma'}}_{\omega,\nu \nu'}$ for both channels are obtained from the single-impurity problem.}

{Further, performing} the same steps as in Refs. \cite{My_BS,MyEDMFT}, we find the irreducible susceptibility $\phi_q$ fulfilling the matrix relation
\begin{equation}
\hat{\chi}_q=(\hat{I} - \hat{\phi_q} \hat{U})^{-1}\hat{\phi_q} \label{phi}
\end{equation}
in the form
\begin{eqnarray}
\hat{\phi_q}  &=&\hat{\chi}_q ^{0}\hat{\gamma_q} +\hat{X}_q, \label{phi1}
\end{eqnarray}
where $\hat{X}_{q}^{\mathfrak{m}\mathfrak{m}'}=\sum_{\nu \notin B}\chi _{q,\nu}^{0,\mathfrak{m}\mathfrak{m}'}$ accounts for the contribution of frequencies beyond the considered frequency box $B$, 
\begin{align}
    \hat{U} = \left(
\begin{matrix}
  0 & 0 & 0 & -U \\
  0 & U & 0 & 0 \\
  0 & 0 & U & 0 \\
  -U & 0 & 0 & 0
\end{matrix} \right),
\label{BareU}
\end{align}
and $\hat{I}$ is the 4$\times$4 identity matrix. The triangular vertex has the form
\begin{align}
    \gamma_q^{\mu \mathfrak{m}'} = \sum\limits_{\nu' \in B} \left[ \hat{I} {\delta}_{\nu\nu'} - ({\hat \Phi}_{\omega, \nu \nu'} - {\widetilde{U}}) \chi^{0}_{ q,\nu} \right]_{\mu \mu'} ^{-1},\label{gamma}
\end{align}
where ${\widetilde{U}}=(\hat{I}-\hat{U} \hat{X}_q)^{-1}\hat{U}$.  
Equations (\ref{phi1}) and (\ref{gamma}) are used in numerical calculations of the irreducible susceptibility $\phi_q$.

In {the} case {of} the wave vector $\mathbf{Q}=(\pi - \delta,\pi)$ it can be shown from the representation of {the} nonlocal Green's function (\ref{GFss1}) that full susceptibility satisfies the symmetry relations
\begin{align}
    {\chi}^{\mathfrak{m}\mathfrak{n}}_{-q_x,  q_y, \omega} &= \eta_\mathfrak{m} \eta_\mathfrak{n} {\chi}^{\mathfrak{m}\mathfrak{n}}_{q_x, q_y, \omega}, \label{chi_imcom_sym_1} \\
    {\chi}^{\mathfrak{m}\mathfrak{n}}_{ q_x, -q_y, \omega} &= {\chi}^{\mathfrak{m}\mathfrak{n}}_{q_x, q_y, \omega}, \label{chi_imcom_sym_2}
\end{align}
where $\eta_{\uparrow \uparrow} = \eta_{\downarrow \downarrow} = 1$ and $\eta_{\uparrow \downarrow} = \eta_{\downarrow \uparrow} = -1$.
In {the} commensurate case $\mathbf{Q}=(\pi,\pi)$, in addition to symmetry relations (\ref{chi_imcom_sym_1}) {and} (\ref{chi_imcom_sym_2}), $\hat{\chi}_q$ satisfies
\begin{align}
    {\chi}^{\mathfrak{m}\mathfrak{n}}_{ -q_x, q_y, \omega} &= {\chi}^{\mathfrak{m}\mathfrak{n}}_{q_x, q_y, \omega}. \label{chi_imcom_sym_3}
\end{align}

From the symmetry relations (\ref{chi_imcom_sym_1}) and (\ref{chi_imcom_sym_3}) it immediately follows that in {the} commensurate case ${\bf Q}=(\pi,\pi)$ the longitudinal and transverse channels are decoupled from  each other:
\begin{align}
    {\chi}^{\mathfrak{\sigma\sigma}\mathfrak{\sigma,-\sigma}} = 
    {\chi}^{\mathfrak{\sigma\sigma}\mathfrak{,-\sigma,\sigma}} 
    ={\chi}^{\mathfrak{\sigma,-\sigma,}\mathfrak{\sigma\sigma}} = 
    {\chi}^{\mathfrak{-\sigma,\sigma}\mathfrak{\sigma\sigma}} 
    = 0.
\end{align}
This decoupling was also noted in Refs. \cite{BS_Toschi,SDW2}.

{To solve} the DMFT impurity problem and calculate local vertices we have used the CT-QMC impurity solver, implemented in the iQIST software package \cite{iQIST}. To obtain static and dynamic properties we had to complete {an} analytical continuation of our results from {the} imaginary frequency axis to {the} real frequency axis. For this purpose we used {the} $ana\_cont$ Python package \cite{ana_cont}.



\section{Results} 

We consider $t'=0.15t$, which is {a} typical value for La$_{2-x}$Sr$_x$CuO$_4$, and set the temperature $T=0.1t$. We calculate the susceptibilities $ \chi^{ab}(\textbf{q},\omega)\equiv\sum_{{\mathfrak m}{\mathfrak m}'} \sigma^a_\mathfrak{m}\chi^{{\mathfrak m}{\mathfrak m}'}_q \sigma^b_{{\mathfrak m}'}$, where $a,b=x,y,z$.
The initial wave vector ${\bf Q}$ 
is obtained {by} searching {the} minimum of the minimal eigenvalue of the matrix ${\hat I}-{\hat \phi}_{\bf Q} {\hat U}$ at $\omega=0$ 
in the antiferromagnetic phase and then determined self-consistently to provide the minimum of the minimal eigenvalue of the same matrix 
in the incommensurate case.  

\begin{figure}[b]
\includegraphics[scale=0.7]{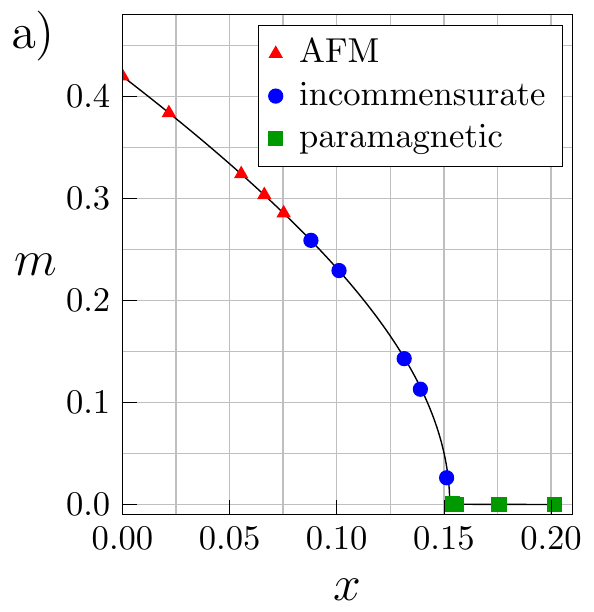}
\includegraphics[scale=0.7]{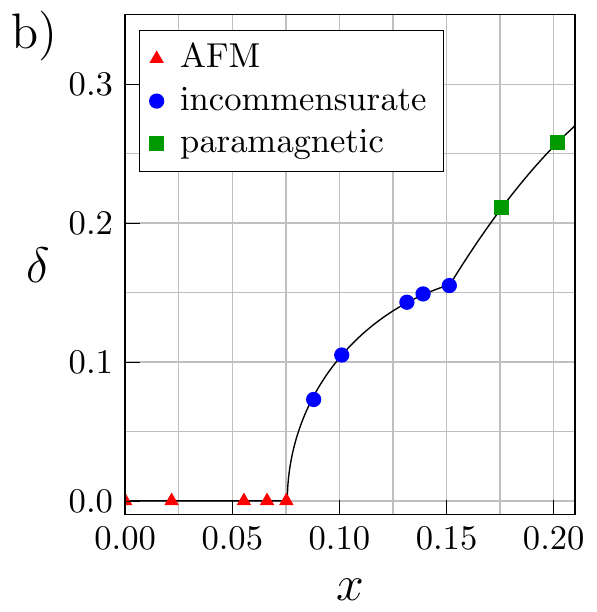}
\caption{(Color online). Dependence of (a) staggered magnetization $m=\langle S^-_{\bf Q}\rangle$ and (b) {the} incommensurability parameter $\delta$ of {the} magnetic order on the hole doping level $x$ at $U/t=7.5$. Triangles denote {the} antiferromagnetic state, circles indicate states with incommensurate magnetic order, and squares represent {the} paramagnetic state. 
}
\label{fig:magnetization}
\end{figure}

\begin{figure}[b]
\includegraphics[width=0.95\linewidth]{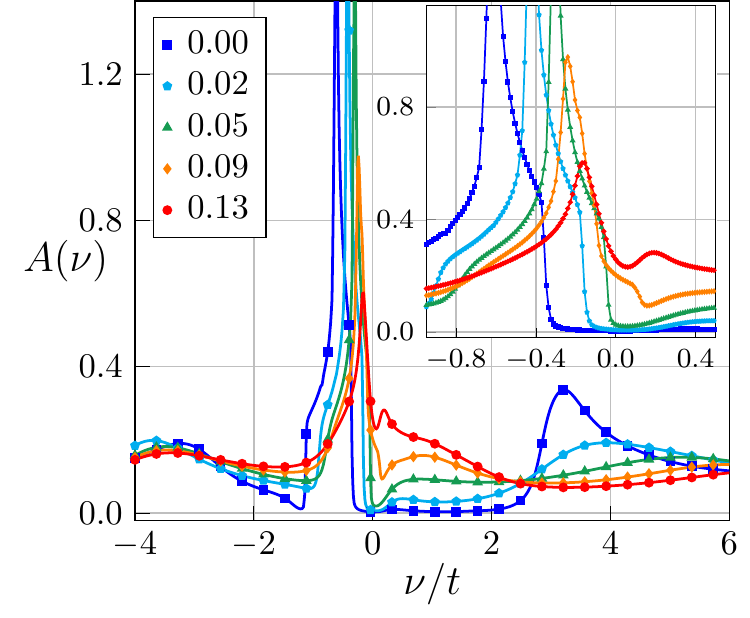}
\caption{(Color online).  Electron {local} spectral functions $A(\nu)$ at $U/t =7.5$ for various values of doping levels $x$ indicated in the legend. The inset shows the low-energy region, containing quasiparticle peaks and the spectral gaps. 
}
\label{fig:spectral_functions}
\end{figure}

\begin{figure}[t]
\vspace{0.2cm}
\includegraphics[width=0.48\linewidth]{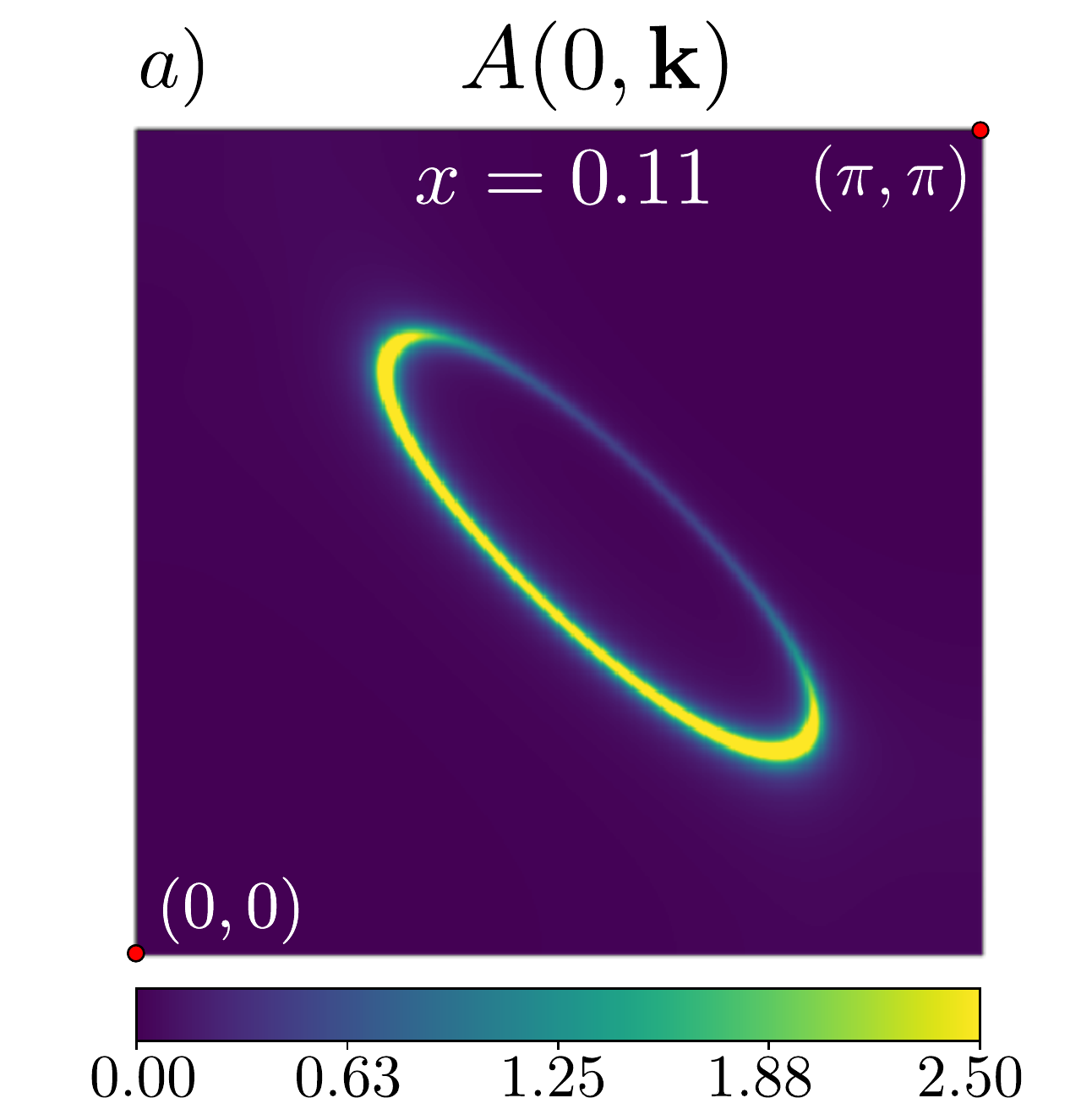}    
\includegraphics[width=0.48\linewidth]{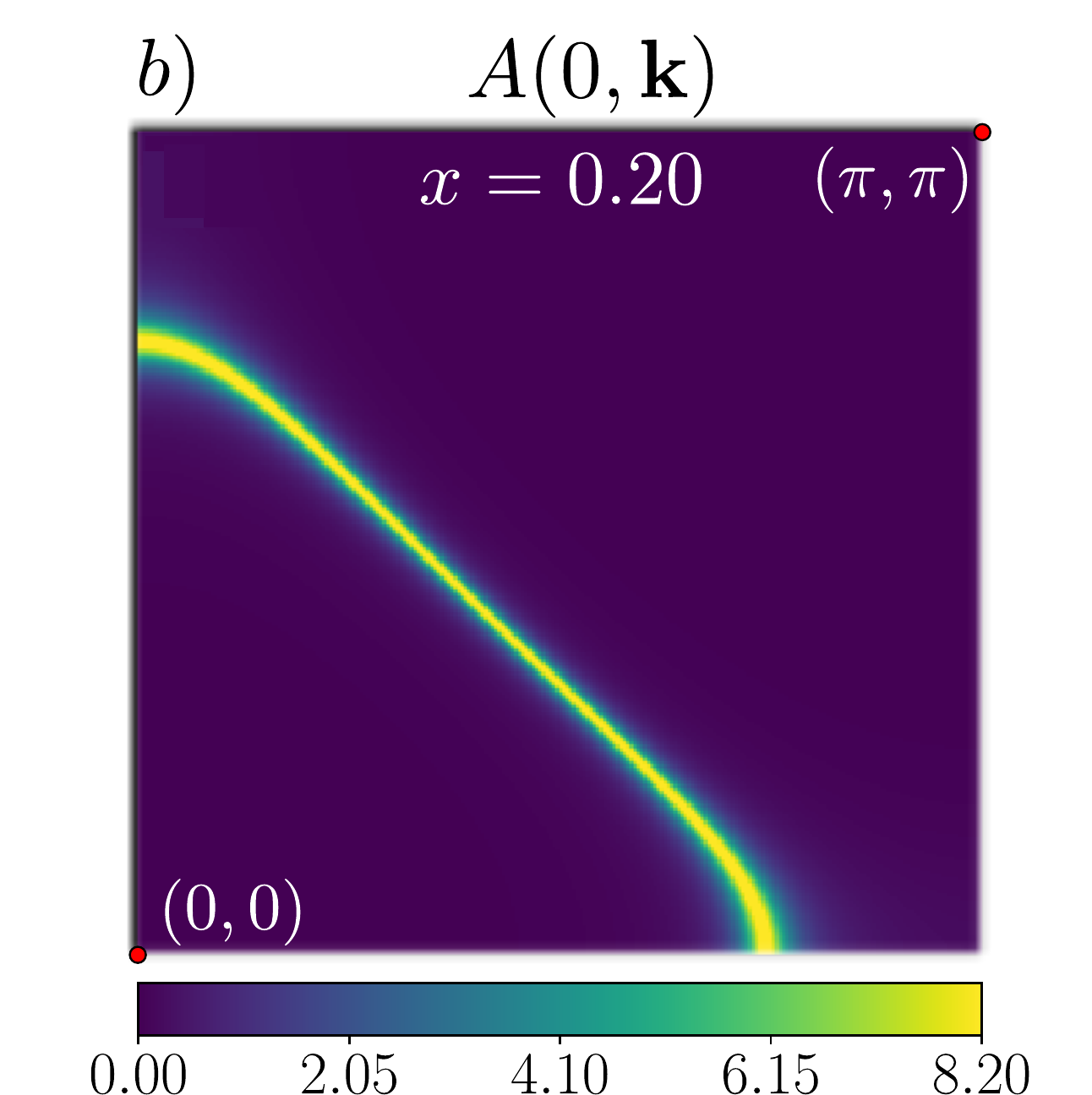}    
\caption{(Color online). {The zero-energy spectral functions}  {showing the positions of the Fermi surface} in the upper right  quadrant of the Brillouin zone at $U/t=7.5$. (a) The state with incommensurate magnetic ordering at hole doping level $x = 0.11$.
(b) Paramagnetic state at hole doping level $x = 0.20$. 
}
\label{fig:fermi_surfaces}
\end{figure}

In all calculations we find that the dominant wave vector ${\bf Q} = (\pi - \delta, \pi)$, in agreement with earlier slave-boson studies \cite{SBIncomm} and similar to the wave vector of short-range magnetic order, observed in La$_{2-x}$Sr$_x$CuO$_4$ \cite{ExpQ1,ExpQ2,ExpQ3,StaticOrd,Plakida}. We note that the mean-field approach \cite{SDW3,SDWOur} yields the dominating phase ${\bf Q}=(Q,Q)$ at {a} sufficiently strong interaction $U$ of the order of the bandwidth, which seems to be the drawback of neglecting correlations in this approach. 

 In Fig. \ref{fig:magnetization}(a) we plot the dependence of staggered magnetization
 $m=\langle {\hat S}^-_{\mathbf Q}\rangle$, where $\hat{S}^{\pm}=\hat{S}^z \pm i \hat{S}^x$,
 on the hole doping $x=1-n$ at $U/t = 7.5$. Staggered magnetization shows mean-field-like behavior
$m \propto \sqrt{x_c - x}$ near the critical value of hole doping level $x_c \approx 0.15$. Although the sublattice magnetization is continuous with doping, there are two different types of states with $x < x_c$. At low doping ($x < x^{(Q)}_c$) we find antiferromagnetic magnetic order [marked by triangles in Fig.  \ref{fig:magnetization}(a)]. 
At intermediate values of {the} hole doping level we obtain incommensurate magnetic order: {the} spin density wave [circles in Fig. \ref{fig:magnetization}(a)]. The dependence of the degree of incommensurability $\delta$ of {the} magnetic order on the doping level $x$ is presented in Fig. \ref{fig:magnetization}(b). Close to {the} commensurate-incommensurate transition this dependence is quite similar to the one, observed for the short-range magnetic order in La$_{2-x}$Sr$_x$CuO$_4$ \cite{ExpQn1,ExpQn2}. 

Using analytical continuation of the {self-energy $\Sigma_\sigma(\nu)$, which takes into account the constant and $1/\nu$ asymptotics of the self-energy at large frequencies \cite{ana_cont,AsymSigma}}, 
we obtain the electron spectral functions $A(\textbf{k},\nu)=- {\rm Im}\mathfrak{G}_{\textbf{k} \sigma}(\nu)/\pi$, as well as the local spectral function $A(\nu)=\sum_{\mathbf k} A(\textbf{k},\nu)$.
In Fig. \ref{fig:spectral_functions} we show $A(\nu)$ for various doping levels. The quasiparticle peak is shifted below the Fermi level and corresponds to the hole states in {the} antiferromagnetic or spiral spin density wave background. At very low doping we find the gap in the spectrum at the Fermi level, {which is present because of the redistribution of the spectral weight between the Hubbard (Slater) subbands, as well as 
thermal activation of the electron-hole pairs (see more details in {the} Appendix).} The gap is continuously filled with {a} further increase of the doping. {Continuous increase of the spectral weight at the Fermi level with doping agrees with the experimental data for La$_{2-x}$Sr$_x$CuO$_4$ \cite{AIPES}. } 

\begin{figure}[b]
\includegraphics[width=1.0\linewidth]{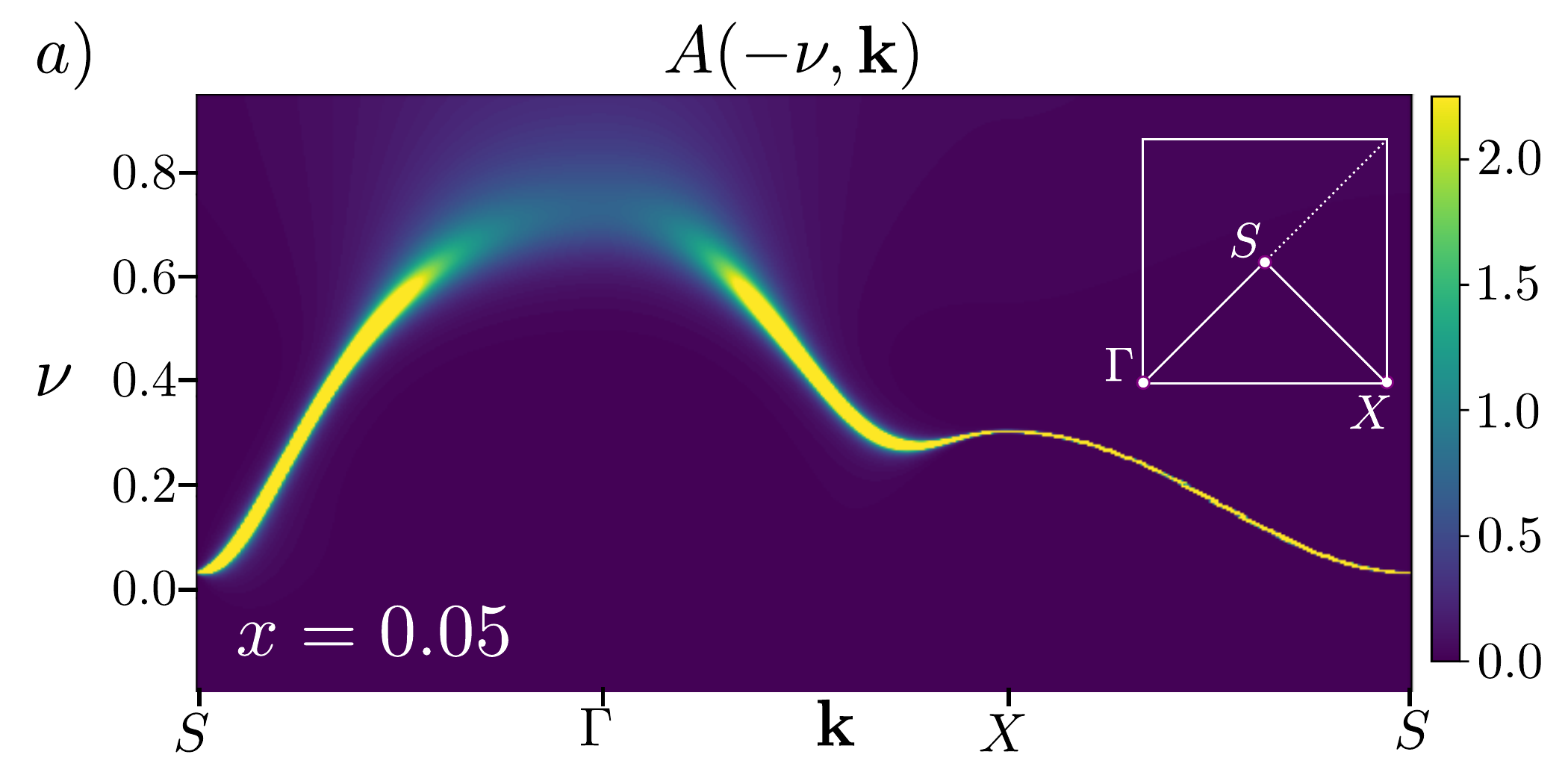}
\includegraphics[width=1.0\linewidth]{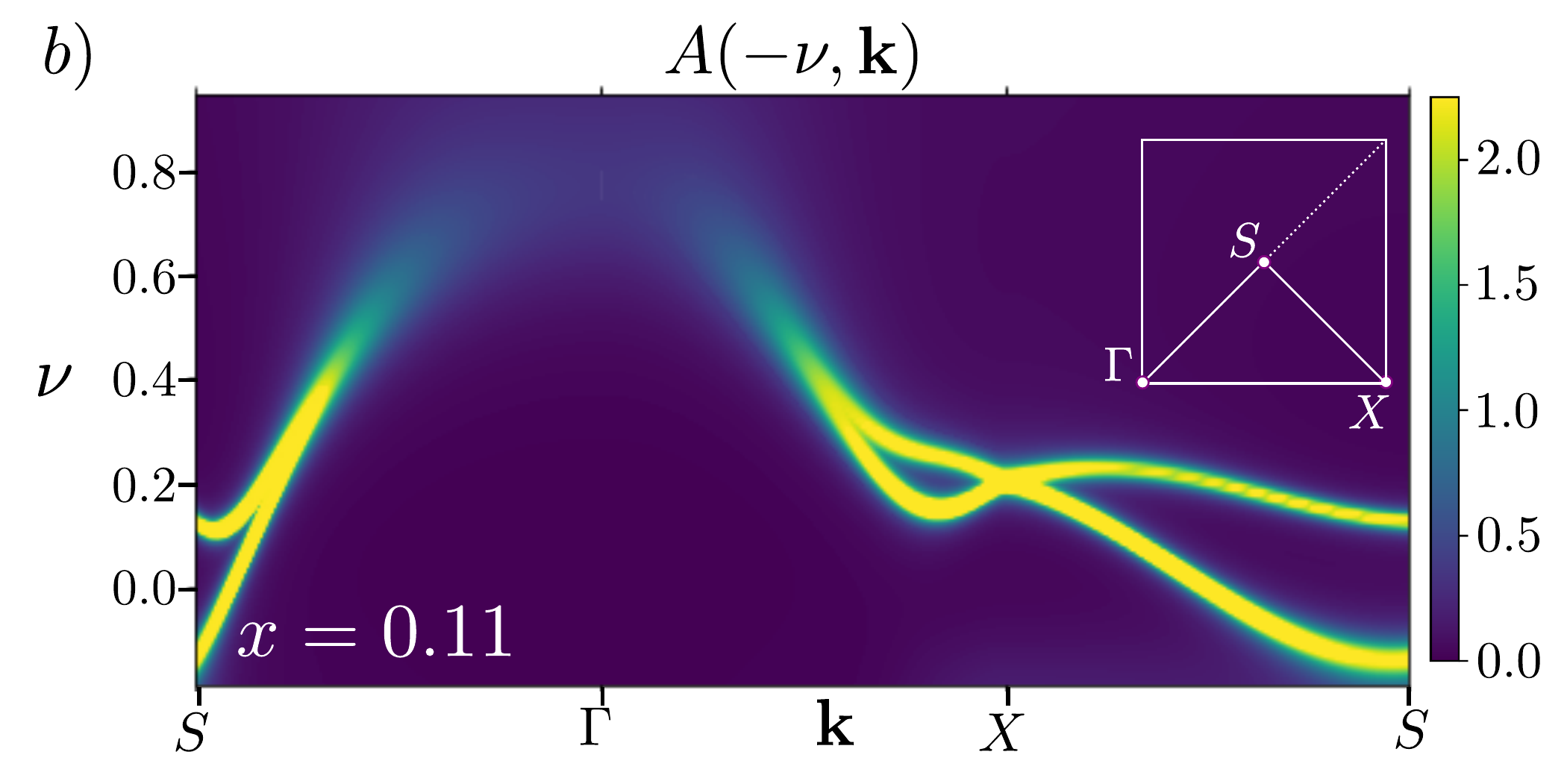}
\caption{(Color online). {The frequency and momentum dependence of the spectral functions at negative energy, showing} the dispersion relation of holes at $U/t=7.5$ in (a) {the} antiferromagnetic state {at} doping $x = 0.05$, and (b) {the} incommensurate state {at} doping $x=0.11$. The Fermi level corresponds to zero energy. Positions of high-symmetry points in the upper right quadrant of the first Brillouin zone are shown in the inset. 
}
\label{fig:dispersion_inc}
\end{figure}

\begin{figure*}[t]
\includegraphics{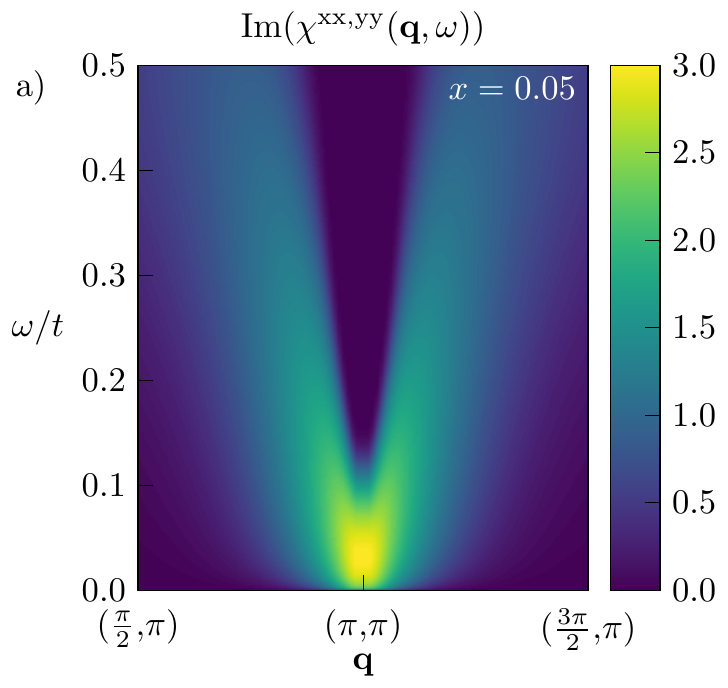}
\includegraphics{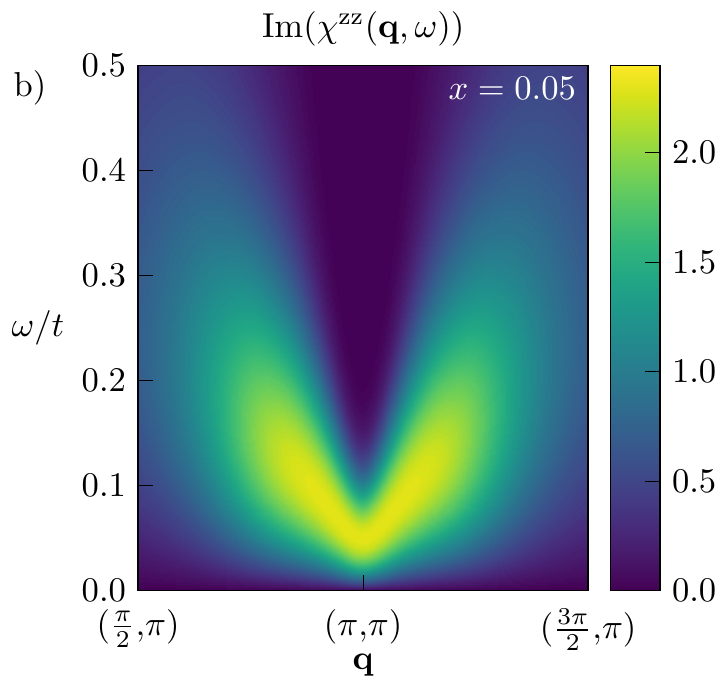}
\caption{(Color online). Imaginary part of (a) the transverse $\chi^{xx,yy}$ (logarithmic color scale) and (b) longitudinal $\chi^{zz}$ (linear color scale) dynamic magnetic susceptibilities as a function of the energy $\omega$ and the vector $\mathbf{q}$ along the path $({\pi}/{2},\pi) \rightarrow ({3\pi}/{2},\pi)$ for {the} antiferromagnetic state at hole doping $x = 0.05$ {and} $U/t=7.5$.
}
\label{fig:chid_zz_afm}
\end{figure*}
\begin{figure*}[t]
\includegraphics{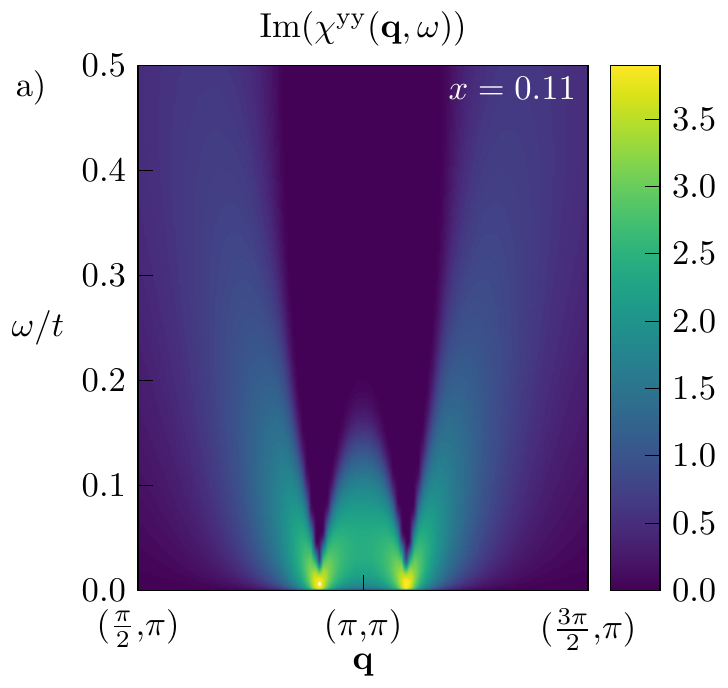}
\includegraphics{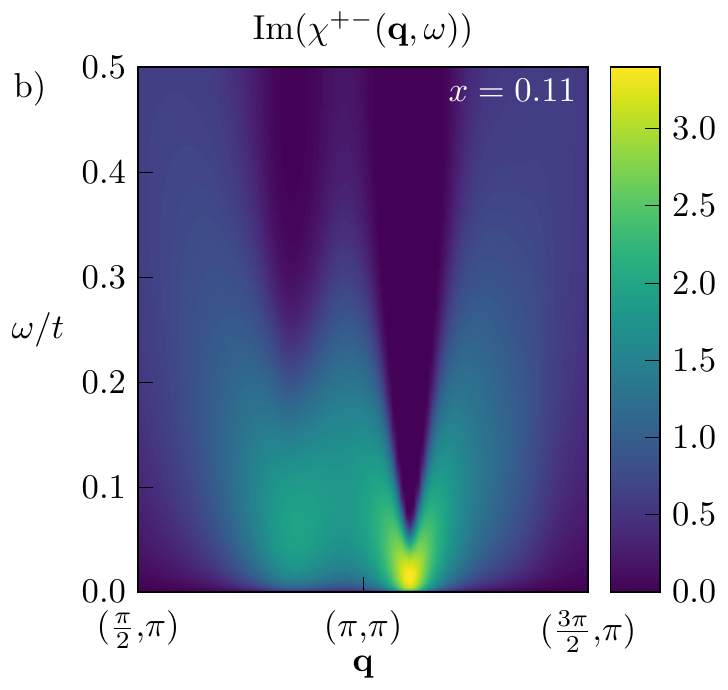}
\caption{(Color online). Imaginary part of (a) the out-of-plane $\chi^{yy}$ and (b) the in-plane $\chi^{+-}$ dynamic magnetic susceptibilities (logarithmic color scale) as a function of the energy $\omega$ and momentum $\mathbf{q}$ along the path $({\pi}/{2},\pi) \rightarrow ({3\pi}/{2},\pi)$ for hole doping $x = 0.11$ (corresponding  incommensurability $\delta = 0.10$) {and} $U/t=7.5$.
}
\label{fig:chi011}
\end{figure*}

In Fig. \ref{fig:fermi_surfaces} we show the {contour plots of the zero energy $A({\mathbf k},\nu=0)$ spectral functions, {whose} maxima show the positions of the} Fermi surfaces. 
{At small hole doping the Fermi pockets are absent since the Fermi level lies in the gap of the spectral function.}
In the incommensurate state [see Fig. \ref{fig:fermi_surfaces}(a) for hole doping level $x = 0.11$] the Fermi surface consists of hole pockets. It is spin independent and symmetric under each of the transformations $k_x \rightarrow - k_x$ and $k_y \rightarrow - k_y$. At the same time, the Fermi surface for {the} paramagnetic state [see Fig. \ref{fig:fermi_surfaces}(b) for the hole doping level $x = 0.20$] is connected and has the standard shape.


\begin{figure*}[t]
\includegraphics{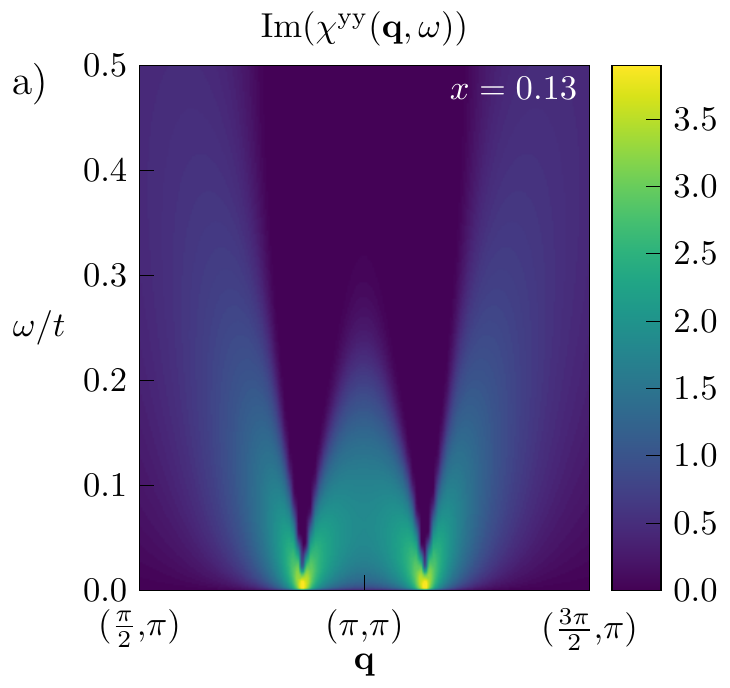}
\includegraphics{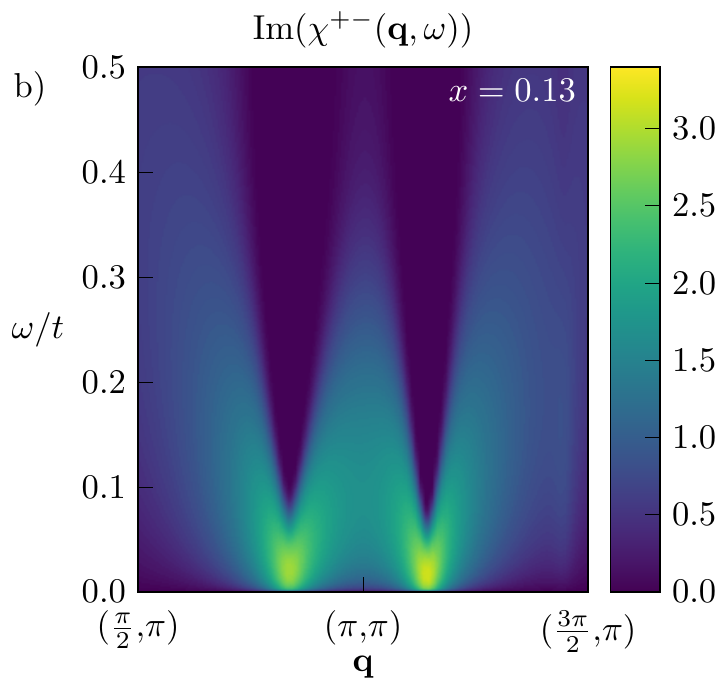}
\caption{(Color online). The same as Fig. \ref{fig:chi011} for hole doping $x = 0.13$ (corresponding  incommensurability $\delta = 0.13$).
}
\label{fig:chic_+-_013}
\end{figure*}
\begin{figure*}[t]
\includegraphics{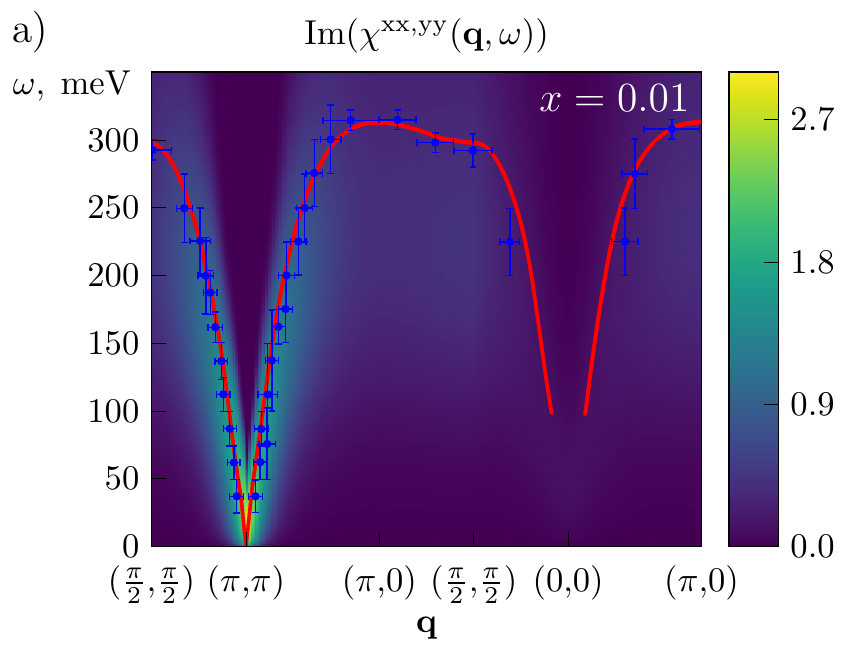}
\includegraphics{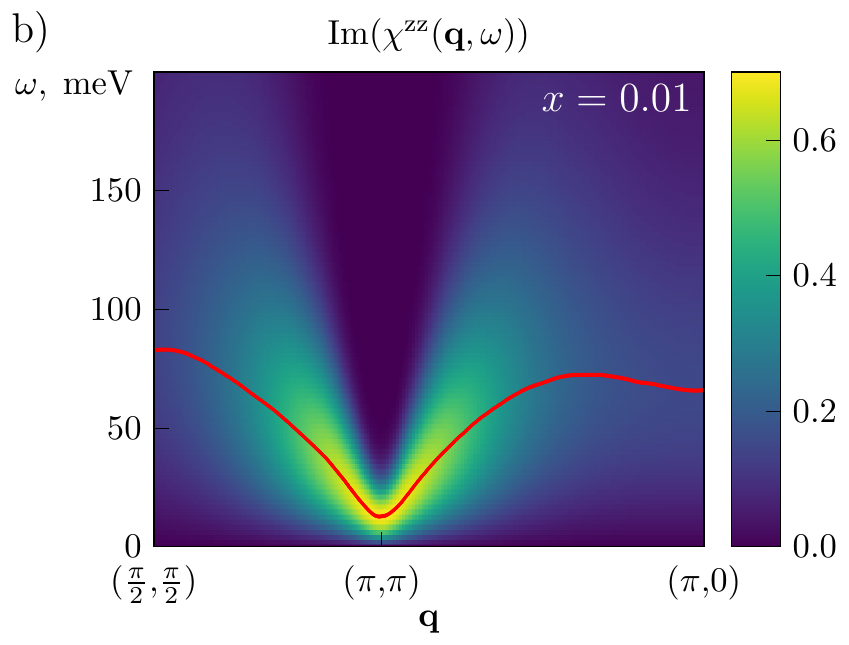}
\caption{(Color online). Imaginary part of (a) the transverse $\chi^{xx,yy}$ (logarithmic color scale) and (b) longitudinal $\chi^{zz}$ (linear color scale) dynamic magnetic susceptibilities as a function of energy $\omega$ and the vector $\mathbf{q}$ along symmetric directions for {the} commensurate antiferromagnetic state near half filling {at hole doping level $x=0.01$}. The interaction $U/t=10$ {and} $t = {\SI{425}{\milli\electronvolt}}$. The maxima of the imaginary parts are shown by {the} red line, except {in} the vicinity of the point ${\bf q}=0$, where the intensity is very low and not captured by analytical continuation. The experimental data for the spin-wave dispersion {at half filling}, taken from Ref. \cite{MagnDisp}, {are} represented in the left plot by blue points with error bars.}
\label{fig:chid_+-}
\end{figure*}

In Fig. \ref{fig:dispersion_inc} we show the hole dispersion, obtained from the contour plots of hole spectral functions $A(\textbf{k},-\nu)$, for {the} antiferromagnetic state (doping $x = 0.05$) and {the} incommensurate state (doping $x=0.11$). {Hole} dispersion calculated in the antiferromagnetic state qualitatively coincides with the results obtained for {the} $t-J$ model (see, e.g., Refs. \cite{tJ1,tJ2}). However, in contrast to earlier studies, in the antiferromagnetic state it does not cross the Fermi level. {Low} spectral weight at the Fermi level, which is seen in Fig. \ref{fig:spectral_functions}, appears in this case from the accumulation of the tails of the peaks in the $k$-resolved spectral functions. At the same time, two different hole dispersion branches are present in {the} incommensurate case, which originate from {the} ${\mathbf k}\pm {\mathbf Q}/2$ contributions in Eq. (\ref{Gcc}). The lower dispersion branch in this case forms the Fermi surface pocket.  When the system {changes} to {the} antiferromagnetic phase, the two modes merge. We have verified that in both commensurate and incommensurate cases the main contribution to the dispersion comes from the first term in the second line of Eq. (\ref{Gcc}) with $\sigma'=\uparrow$, i.e., the diagonal spin-majority states.

The local vertex approximation and analytical continuation of bosonic quantities allow us to calculate different components of the dynamic susceptibility $\chi(\textbf{q},\omega)\equiv\chi_q$. {Its matrix-valued} nature in our formalism {provides} an opportunity to investigate different types of magnetic excitations, which are coupled to each other for the incommensurate order, as was emphasized in Ref. \cite{SDW3}.

In Fig. \ref{fig:chid_zz_afm} we show the corresponding susceptibilities in the doped commensurate case for $x=0.05$. As discussed in Sec. \ref{Rel2P}, in the state with the commensurate antiferromagnetic order the ``longitudinal'' excitations, determined by the maximum of the susceptibility ${\rm Im}\left[\chi^{zz}(\textbf{q},\omega)\right]$ are decoupled from the transverse ones, determined by ${\rm Im}\left[\chi^{xx,yy}(\textbf{q},\omega)\right]$. From the $\chi^{xx,yy}(\textbf{q},\omega)$ susceptibilities we obtain massless Goldstone modes, which are related to the spontaneous symmetry breaking. Their spectrum broadens rapidly with the deviation of the wave vector $\mathbf{q}$ from $\mathbf{Q}=(\pi,\pi)$.
{We do not find evidence {of} 
negative modes in the magnon's spectrum. 
This shows that {the} considered ${\mathbf Q}=(\pi,\pi-\delta)$ order is stable.}
The longitudinal excitations, obtained from  ${\rm Im}\chi^{zz}(\textbf{q},\omega)$, possess a gap of the order {of} $0.1t$. Similar to the transverse channel, the longitudinal excitations become less coherent with {the} deviation of the {wave vector} $\mathbf{q}$ from $\mathbf{Q}$.


In the incommensurate case the longitudinal and transverse excitations are coupled. Accordingly, we analyze the in-plane excitations, determined by ${\rm Im}(\chi^{+-}\left[\textbf{q},\omega)\right]$ susceptibility of $\hat{S}^{\pm}$ spin components, and the out-of-plane component, determined by ${\rm Im}\left[\chi^{yy}(\textbf{q},\omega)\right]$.
In Fig. \ref{fig:chi011} we show the corresponding susceptibilities for {the} incommensurate state far from transitions to paramagnetic and antiferromagnetic states. In ${\rm Im}\left[\chi^{yy}(\textbf{q},\omega)\right]$ we find two Goldstone modes at wave vectors ${\mathbf q}=\pm \mathbf{Q}$. They merge at ${\mathbf q}=(\pi,\pi)$ when {the} order {changes} to {the} antiferromagnetic one.
In the in-plane component ${\rm Im}\left[\chi^{+-}(\textbf{q},\omega)\right]$ there is also {a} Goldstone mode at ${\mathbf q}=-{\mathbf Q}$, corresponding to transverse excitations, while at ${\mathbf q}={\mathbf Q}$ we observe {a} longitudinal (Higgs) mode of weaker intensity. As in the commensurate case, this mode is gapped. Upon approaching the paramagnetic phase (see Fig. \ref{fig:chic_+-_013}) the longitudinal mode softens, and the corresponding gap disappears at the spin-density wave to paramagnet transition.
{{Like in} the commensurate case, no evidence of instability of the 
considered long-range order is obtained, such that this order appears to be stable.}

We have verified that the presented results remain {qualitatively} unchanged for larger values of $U/t$, except that the incommensurate region of {doping levels} becomes narrower with {the} increase of $U/t$ at fixed temperature $T$. The considered value $U/t=7.5$
corresponds to the metallic state at half filling in the absence of long-range antiferromagnetic order, and it is also somewhat smaller than typically considered for cuprate high-$T_c$ compounds. To compare the obtained magnon dispersions with the experimental data for cuprates, in Fig. \ref{fig:chid_+-} we show the magnon dispersion for $U=10t$. {We consider {an} antiferromagnetic state at small doping $x=0.01$ to avoid difficulties {with} analytic continuation of dynamic susceptibilities at half filling. The obtained dispersion is compared to the experimental data of Ref. \cite{MagnDisp} at half filling}. {We} can see that for the hopping {$t=425$~meV} the obtained maxima of ${\rm Im}\chi^{xx,yy}({\bf q},\omega)$ compare well {with} the experimental dispersion, including the $(\pi,0)$-$(\pi/2,\pi/2)$ part, {whose} deviation from the flat dispersion in the linear spin-wave analysis of the Heisenberg model was previously attributed to the ring exchange (see, e.g., Refs. \onlinecite{MagnDisp,KK}). Taking into account the renormalization factor of the spin-wave dispersion of the $S=1/2$ two-dimensional Heisenberg model \cite{SW,OurSW} $\gamma=1.157$, which originates from the magnon interaction and not accounted {for} by the considered ladder approximation, we find the bare hopping, capable {of describing} the magnon dispersion $t_{\rm bare}=t/\gamma\simeq{370}$~meV, in good agreement with the estimate $J_{\rm bare}U/(4t)\simeq t_{\rm bare}$, considering $J_{\rm bare}\simeq {152}$~meV \onlinecite{KK}. The maximum of the imaginary part of the longitudinal susceptibility shows longitudinal excitation with a rather small gap $\sim 10$~meV.

\section{Conclusion} 
In the present paper we considered {the} calculation of sublattice magnetization, {the} incommensurability parameter, {hole} dispersion, and {the} dynamic magnetic susceptibilities of the square lattice Hubbard model with nearest- and next-nearest-neighbor hopping in the antiferromagnetic and incommensurate cases within the dynamic mean-field theory. At small doping we obtained {an} antiferromagnetic insulating state with hole dispersion, which agrees qualitatively with that obtained previously for the $t$-$J$ model.
The transverse magnetic susceptibility possesses a Goldstone mode, while the longitudinal excitations are characterized by a small gap. 

At larger doping we {found an} incommensurate {spiral ${\mathbf Q}=(\pi,\pi-\delta)$} order with two branches of hole dispersions and Fermi surfaces, having the shape of hole pockets. The out-of-plane magnetic susceptibility possesses Goldstone modes at the incommensurate wave vectors, while the in-plane susceptibility shows both, gapless Goldstone spin-wave excitations and the longitudinal gapped excitation. {The obtained long-range order {was} found to be stable from the susceptibility analysis.}

We also showed that close to half filling the experimentally observed magnon dispersion can be reproduced with reasonable values of hopping $t$ and interaction $U$. 
The obtained doping evolution of the Fermi surfaces and the incommensurability parameter $\delta$ 
are quite similar to those for high-$T_c$ cuprate superconductors, which stresses once more {a} possible magnetic origin of the pseudogap. Although the considered magnetic state is longrange ordered, this order can be "hidden", e.g., within the gauge theory of fluctuating spin-density wave order discussed recently in Refs. \cite{Sachdev1,Sachdev2,PseudogapMag3}. {The} extension of the proposed approach to {the} paramagnetic phase within these gauge theories {therefore} represents {a} promising direction of future research. This may also concern other types of lattices, including compounds with {a} frustrated triangular lattice, which are candidates for the spin-liquid phase \cite{SL1,SL2,SL3,SL4}.


\section*{Acknowledgements}
The authors acknowledge the financial support from the BASIS Foundation (Grant No. 21-1-1-9-1) and the Ministry of Science and Higher Education of the Russian Federation (Agreement No. 075-15-2021-606). A. A. K. also acknowledges the financial support within the theme ``Quant" 122021000038-7 of {the} Ministry of Science and Higher Education of the Russian Federation.

\appendix*

\section{Relation to the static mean-field approach}

The static mean-field treatment in the considered approach is conveniently formulated by considering the (frequency-independent) mean-field approximation for the self-energy of the impurity Anderson model $\Sigma_{\sigma} = U n_{-\sigma}$. The corresponding occupation numbers are obtained from the equation 
    \begin{equation}
        n_\sigma = T \sum_{\nu} \frac{1}{\zeta_{\sigma}^{-1}(i\nu)-U n_{-\sigma}}
    \end{equation}
Exploiting the self-consistency equation (\ref{self_con_eqv}) we obtain
    \begin{equation}
        T  \sum_{\mathbf{k},\nu} G_{\mathbf{k}}^{\sigma\sigma}(i\nu) = n_{\sigma}
        \label{mean_field_sc}
    \end{equation}
Defining the mean-field order parameter
\begin{equation}
\Delta = \frac{\Sigma_{\downarrow}-\Sigma_{\uparrow}}{2}={U m},
\end{equation}
where the local magnetization $m=(n_\uparrow-n_\downarrow)/2$, and taking into account the form of the Green's function (\ref{Gss}), the summation over frequencies in Eq. (\ref{mean_field_sc}) can be performed analytically and yields the equation
\begin{equation}
        \frac{1}{U} = \sum_{\mathbf{k}} \frac{f(E_{v}({\mathbf k}))-f(E_c({\mathbf k}))}{2 E_{-}(\mathbf{k})},
        \label{GapMF}
    \end{equation}
where 
    \begin{align}
    E_{c,v}({\mathbf k})&=\epsilon_{+}(\mathbf{k})\pm E_{-}(\mathbf{k})-{\tilde\mu},\\
E_{-}(\mathbf{k}) &= \sqrt{\epsilon_{-}^2({\mathbf k})+\Delta^2},\\
        \epsilon_{\pm}(\mathbf{k}) &= \frac{1}{2}\left(\epsilon_{\mathbf{k-Q}/2} \pm \epsilon_{\mathbf{k+Q}/2}\right), 
    \end{align}
$f(\varepsilon) = \left[\exp(\varepsilon/T)+1\right]^{-1}$ is the Fermi function, and ${\tilde \mu}=\mu-(\Sigma_\uparrow+\Sigma_\downarrow)/2$.  Equation (\ref{GapMF}) allows one to determine the order parameter $\Delta$ (the chemical potential $\tilde \mu$ is fixed by the total number of particles) and coincides with the corresponding mean-field equation of Refs. \cite{SDW1,SDW2,SDWIc1,SDWIc2,SDWIc3,SDW3,SDW4,SDWOur,BM}.

Analogously, the ladder summation of diagrams for susceptibilities, considered in Refs. \cite{SDW1,SDW2,SDWIc2,SDW3,SDWIc3,SDW4,BM}, can be reproduced for {an} infinite (finite) frequency box considering the bare two-particle irreducible vertices ${\hat \Phi}_{\omega,\nu\nu'}=\hat{U}$ (${\tilde U}$), where the matrix $\hat{U}$ (${\tilde U}$) is defined in Eq. (\ref{BareU}) [after Eq. (\ref{gamma})]. This also implies the triangular vertex $\gamma_q^{\mu {\mathfrak m}'}=\delta_{{\mathfrak m}{\mathfrak m}'}$ in the mean-field approximation.

\begin{figure}[t!]
\vspace{.5cm}
\includegraphics[width=0.95\linewidth]{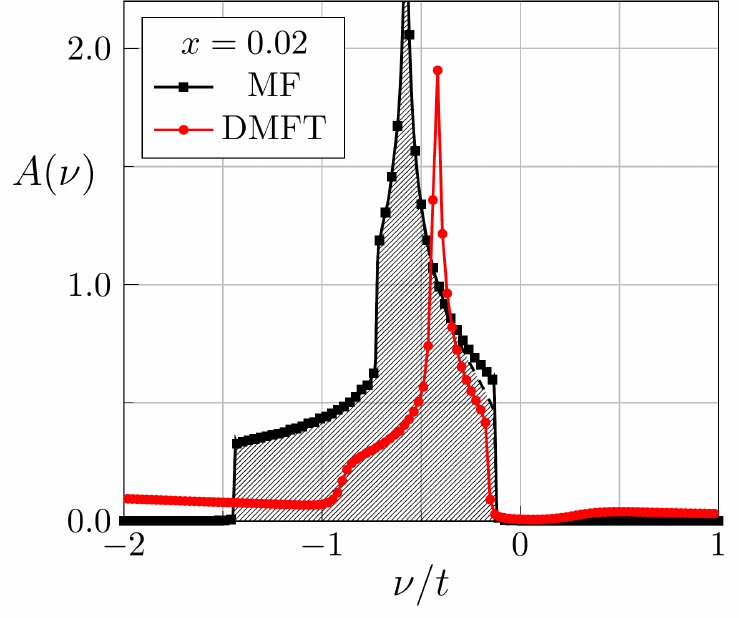}
\includegraphics[width=0.95\linewidth]{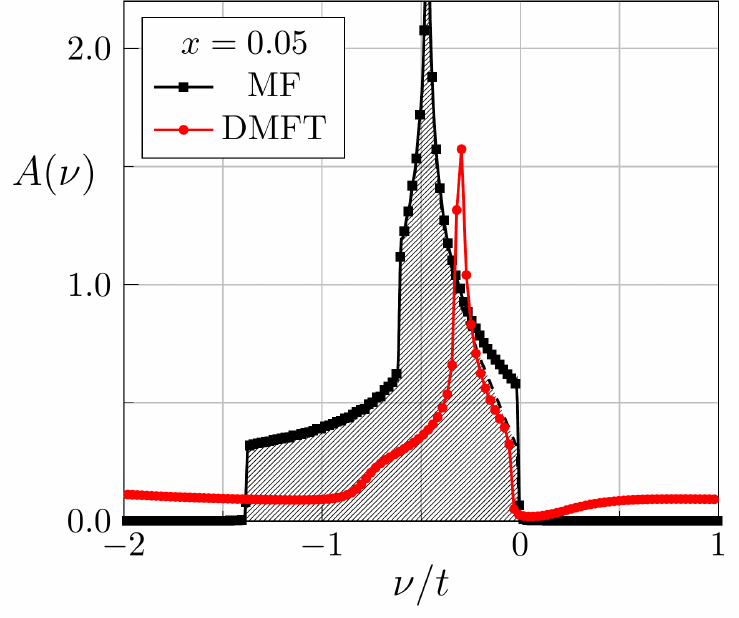}
\caption{(Color online). Electron {local} spectral functions $A(\nu)$ near the Fermi level for $U/t =7.5$, $T=0.1t$, and hole doping levels $x = 0.02$ and $x=0.05$ obtained {using the} mean-field approach (black squares). {The} shaded area under the dashed line, showing $A(\nu)f(\nu)$, represents partial occupancy of electronic states due to thermal smearing. The results of the DMFT approach from Fig. \ref{fig:spectral_functions}  are shown by red circles.
}
\label{fig:spectral_functions_0.02}
\end{figure}

{The density of states of the lower Slater (valence) band $E_v({\mathbf k})$ in the mean-field approach for parameter values $t'=0.15t$, $U=7.5t$, {and} $T=0.1t$, considered in the {main text}, is shown in Fig. \ref{fig:spectral_functions_0.02}. For the purpose of the comparison to {the} DMFT approach we fix the wave vector to ${\mathbf Q}=(\pi,\pi)$.}
{At the considered small doping and finite temperature the Fermi level lies within the spectral gap; thermal activation of holes in the valence band results in {an} occupancy smaller than {$1$}, corresponding to the chosen hole doping level. This effect of thermal smearing is illustrated in Fig. \ref{fig:spectral_functions_0.02} by {the} shaded area under the mean-field spectral function, which represents {the} fraction of occupied states. With increasing hole doping, the Fermi level shifts closer to the valence band. This behavior of {the} spectral function in the mean-field approach is preserved also for small deviations of magnetic order from the N\'eel state.}


While previous mean-field studies \cite{SDW3,SDW4,SDWOur} showed that at the mean-field level the spiral phases are thermodynamically unstable at sufficiently strong interaction and show the regions of negative susceptibilities, as we discussed in the main text, this drawback of the mean-field theory is cured in the dynamical mean-field theory approach. As also discussed in Ref. \cite{vDongen}, the static mean-field solution is not quantitatively correct for the symmetry-broken phases even in the limit of weak coupling. In particular, for the antiferromagnetic case, already at sufficiently small interaction, the DMFT solution differs from that of the static mean-field approach due to {the} {development} of the frequency dependence of the self-energy \cite{ToschiAF}. The same is expected for the incommensurate phases. 

{ The spectral functions of the DMFT approach at small doping, presented in the main text (Fig. \ref{fig:spectral_functions}), show {a} quasiparticle peak {with the} right edge only slightly shifted with respect to that of the lower Slater band of the mean-field theory, but the height of the peak remains smaller than in the mean-field theory due to partial transfer of the spectral weight to the lower Hubbard band at negative energies (see {the} comparison in Fig. \ref{fig:spectral_functions_0.02}). Although at finite temperatures the quasiparticle peak is also not fully filled in {the} DMFT even for the Fermi level position above the peak, in contrast to the mean-field approach, the spectral weight of the quasiparticle peak and lower Hubbard band (which we define as the integral weight  $w=\int_{-\infty}^0 d\nu A(\nu)$)
in {the} DMFT reduces with doping (see Figs. \ref{fig:spectral_functions} and \ref{fig:spectral_functions_0.02}). In particular, for $x=0.02$ and $x=0.05$ we obtain for this weight $w=0.99$ and $w=0.97$, respectively \cite{NotePade}, such that in {the} DMFT the spectral weight is redistributed between the lower and upper Hubbard (Slater) subbands. {Therefore, in contrast to the mean-field approach, the DMFT approach, apart from the trivial thermal smearing, shows {an} additional many-body contribution to the spectral functions, which keeps the Fermi level inside the gap at low doping.}




\begin{thebibliography}{1}
\bibitem{NotePade} {We have verified that very close values are obtained using Pad\'e analytical continuation of the self-energy with {a} fixed constant and $1/\nu$ asymptotics at large frequencies.}
\end{thebibliography}
\end{document}